# Electron energy increase in a laser wakefield accelerator using longitudinally-shaped plasma density profiles

Constantin Aniculaesei[1,#], Vishwa Bandhu Pathak[1], Hyung Taek Kim[1,2,*], Kyung Hwan Oh[1], Byung Ju Yoo[1], Enrico Brunetti[4], Yong Ha Jang[1], Calin Ioan Hojbota[1,3], Junghun Shin[1], Jeong Ho Jeon[1], Seongha Cho[1], Myung Hoon Cho[1], Jae Hee Sung[1,2], Seong Ku Lee[1,2], Björn Manuel Hegelich[1,3] and Chang Hee Nam[1,3]

[1]Center for Relativistic Laser Science, Institute for Basic Science (IBS), Gwangju 61005, Republic of Korea.
[2]Advanced Photonics Research Institute, Gwangju Institute of Science and Technology (GIST), Gwangju 61005, Republic of Korea
[3]Department of Physics and Photon Science, GIST, Gwangju 61005, Republic of Korea
[4]Scottish Universities Physics Alliance, University of Strathclyde, Department of Physics, Glasgow, G4 0NG, United Kingdom
[*]htkim@gist.ac.kr
[#]ca182@ibs.re.kr

*The phase velocity of the wakefield of a laser wakefield accelerator can, theoretically, be manipulated by shaping the longitudinal plasma density profile, thus controlling the parameters of the generated electron beam. We present an experimental method where using a series of shaped longitudinal plasma density profiles we increased the mean electron peak energy more than 50%, from 174.8 ± 1.3 MeV to 262 ± 9.7 MeV and the maximum peak energy from 182.1 MeV to 363.1 MeV. The divergence follows closely the change of mean energy and decreases from 58.95 ± 0.45 mrad to 12.63 ± 1.17 mrad along the horizontal axis and from 35.23 ± 0.27 mrad to 8.26 ± 0.69 mrad along the vertical axis. Particle-in-cell simulations show that a ramp in a plasma density profile can affect the evolution of the wakefield, thus qualitatively confirming the experimental results. The presented method can increase the electron energy for a fixed laser power and at the same time offer an energy tunable source of electrons.*

## INTRODUCTION

A high power laser pulse travelling into a plasma pushes forward and radially the background electrons due to the ponderomotive force[1,2] which is proportional to the laser pulse intensity gradient. The ions left behind are quasi-stationary due to their heavier mass and form a positively charged region. As the laser pulse passes, the background electrons are attracted by the local space-charge field and start to oscillate around the ions at the plasma frequency setting up a plasma wave trailing behind the laser pulse. This kind of wave is called a wakefield and constitutes the essential part of a laser wakefield accelerator (LWFA)[3].

In a highly nonlinear case, called bubble[4] or blowout regime[5] the wakefield takes a spherical shape and the electrons from the back of the bubble, if they have proper momentum, are injected into the bubble and are accelerated by the difference of electric potential existing in the bubble; the potential has a minimum at the back of the bubble and a maximum at the centre of the bubble. Note that the minimum potential corresponds to the maximum accelerating field for the electron. The structure of the electric field inside the wakefield resembles the one found in a conventional radio-frequency particle accelerator, yet, it is at least three orders higher in magnitude (electric fields of the order of 100 GV/m). The possibility of obtaining compact ultra-short[6,7] and ultra-bright electron bunches[8–10] and radiation sources[11–15] explains why the scientific community is so excited about the prospects of LWFA.

In a plasma-based electron accelerator, the acceleration length (AL) is one of the parameters that determine the final electron energy. For a self-guided laser pulse, the AL is limited either by the laser etching (depletion) length [$L_{etch} \approx (\omega_0/\omega_p)^2 c\tau_L$], or, for a non-evolving bubble, by the dephasing length [$L_d = (4/3)(\omega_0^2/\omega_p^2)\sqrt{a_0}\, c/\omega_p$], where $a_0 = eE_0/(m_e c\omega_0)$ is the normalized vector potential of the laser, c is the speed of light in vacuum, and $E_0$, $\omega_0$ and $\tau_L$ are the peak laser electric field, laser central frequency and laser pulse length respectively. Plasma frequency is defined by the formula $\omega_p = 4\pi n_0 e^2/m$ with $n_0$ being plasma density, $m$ is electron rest mass and $e$ is the elementary charge. However, during the laser propagation due to the temporal and spatial self-compression[16], the laser intensity significantly increases which leads to the bubble expansion. The bubble expansion effectively reduces the accelerating field gradient seen by the electron bunch, significantly reduces the dephasing length and, as a consequence, reduces the final energy gain. Even though at the end acceleration the bubble shrinks due to laser absorption, however, it is not sufficient to compensate the energy-loss in dephasing due to bubble expansion. Therefore, to achieve the maximum energy gain from an LWFA, the lock/synchronization of the electrons at the maximum field of the wakefield is required.

A method to synchronize the electron bunch with the wakefield phase has been proposed first by Katsouleas[17] and uses an up-ramp plasma density. In an upward plasma density ramp the phase velocity of the back of the bubble continuously increases (the bubble shrinks), and for a proper plasma density gradient, matches the group velocity of the electron bunch. In the frame of reference of the electron bunch, in these conditions, the back of the bubble is stationary. This process is called phase-locking and it has been theoretically shown[18–22] that it can control the parameters of the electron beam such as energy and emittance/divergence.

In this report, we present a simple, yet powerful, experimental method that uses a series of longitudinally tailored plasma profiles to control the phase between electron bunch and the wakefield thus, consequently, controlling the parameters of the accelerated electron bunch. Although some experimental work has been done[23] to study the influence of the linear density ramps, to the author's knowledge there is no systematic experimental investigation of the phase-locking mechanism of a laser wakefield accelerator in long up-ramp plasma density.

# Experimental setup and methods

**Laser system**

The high power laser with a central wavelength of 810 nm and horizontal polarization is based on chirped pulse amplification technique[24] and is capable of delivering 3.9 J pulses (after compression) with a temporal duration of 27 ± 2 femtoseconds at a repetition frequency of 5 Hz. During the experiment, 1.8 J laser energy per pulse has been used. A spherical mirror with the focal length of 1.5 meters focused down the 65 mm diameter laser beam to a focal spot with a diameter of 42 microns at full-width-at-half-maximum (FWHM) which contained 52% from the total energy. This ensured on target a normalized vector potential of $a_0 = 1.3$.

**Fluid dynamics simulations of the supersonic gas jet**

The fluid dynamics simulations of the gas nozzle used in the experiment have been performed using the commercial code Ansys Fluent[25] that solved the Navier-Stokes equations on a 2-dimensional axisymmetric mesh containing on average $10^6$ quadrilateral cells with various volumes between $10^{-14}$ m$^3$ and $10^{-16}$ m$^3$. A turbulent model, the k-ω shear stress transport model[26], has been used with double precision. *k* represents turbulence kinetic energy and *ω* represents the specific rate of dissipation. The gas inlet has been chosen as pressure inlet and feeds He gas at various pressures. The gas temperature at the inlet is set at 300 K. The outlet was chosen as pressure outlet at $10^{-3}$ mbar pressure and the walls were considered adiabatic and smooth (no asperities of the surface are present). Mesh-independence tests have been run to ensure that the simulation result does not depend on the mesh size. Similar simulations done using the same conditions showed a very small difference, of the order of 5%, between simulation and interferometry results[27] thus validating the use of the model in our simulations. 2-D contours of gas density were extracted from simulations and from these lineouts of gas density profiles were taken at different heights above the exit of the nozzle based on the experimental conditions such as the tilting angle and laser-gas jet interaction point.

**The laser wakefield accelerator**

The experiment has been run at Center for Relativistic Laser Science (CoReLS), Institute for Basic Science (IBS), South Korea and the setup is shown in 3.

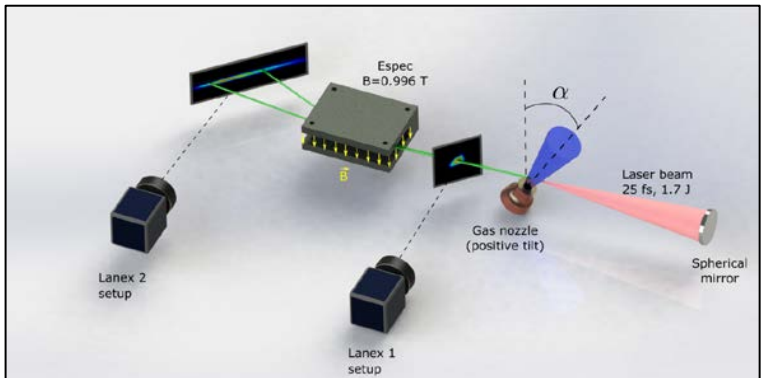

*Figure 1 The setup used for laser wakefield acceleration. It consists of a high power femtosecond laser focused down onto a gas jet tilted at various angles relative to the direction of propagation of the laser. The generated electron beam divergence, relative charge and pointing were obtained from Lanex$^{TM}$ 1 setup which consisted of a Lanex$^{TM}$ screen imaged onto a CCD camera. The electron beam was dispersed horizontally depending on its energy by the magnetic field from the Espec onto a Lanex$^{TM}$ 2 screen. A CCD camera captures the optical emission from the screen and converts it into a false colour image which represented, after calibration, the charge density dispersion as a function of its energy.*

The setup was installed in a vacuum chamber maintained at a pressure of $10^{-5}$ mbar and consisted of a high power laser, focusing optics, gas nozzle and electron beam diagnostics. A Ti:Sapphire 810 nm, 1.8 J, 27 ± 2 fs laser with a diameter of 65 mm is focused onto a supersonic gas jet. The gas jet was generated by a de Laval-type nozzle mounted on a goniometer which can tilt the gas jet at angles of ±45 degrees relative to the direction of propagation of the laser. For the present work, the nozzle has been tilted in the positive direction only (see Fig.3). The gas nozzle has a conical geometry with 1 mm inlet diameter, 3 mm outlet diameter and 6° semi-opening angle. To limit the amount of gas released in the chamber during the operation, the gas nozzle was mounted on a solenoid valve (Parker Series 9) and operated in pulsed mode with an opening time of 3 ms; the laser pulse interacted with the gas jet 1.5 ms after valve opening. The gas density profile was adjusted by tilting the nozzle and afterwards by changing the interaction point relative to the exit of the nozzle (the height) and empirically varying the inlet pressure until an electron beam is obtained. In our setup, the longitudinal direction represents the laser propagation direction and the transverse direction is the horizontal direction. The laser was placed transversely in the centre of the gas nozzle and longitudinally at 1.5 mm from the centre of the nozzle for the entire experiment. The transverse and longitudinal alignment was done with a Top View imaging setup, which consists of a CCD camera (FLIR BFLY-U3-50H5M-C) equipped with a camera lens (Samyang 100mm F2.8 ED UMC MACRO) and a BG39 glass filter to cut the scattered light from the laser. The Top View setup was also used to monitor the optical emission from the plasma channel due to nonlinear Thomson scattering[28,29] and other plasma emission processes[30].

The Lanex[TM] 1 setup consisted of a 6×6 cm $Gd_2O_2S$:Tb screen (Lanex[TM] Back) placed at 390 mm from the gas jet and imaged on a CCD camera (FLIR BFLY-U3-50H5M-C) with a camera lens (Samyang 135mm F2.0 ED UMC), providing information on the transverse electron profile, pointing, divergence and relative charge. The results for each dataset (taken for the same experimental conditions) were averaged and the mean value and the mean standard error was calculated.

The electron spectrometer Espec consisted of a 0.996 T dipole magnet that has an opening with a height, width and length of 8×70×205 mm. The Espec was placed at 430 mm distance from the gas nozzle and dispersed the electrons in the horizontal direction as a function of their energy onto a 430×80 mm $Gd_2O_2S$:Tb screen (Lanex[TM] Back) placed 1125 mm from the gas nozzle. The Lanex[TM] screen was imaged onto a CCD camera (FLIR GS3-U3-50S5M-C) with a lens (Samyang 35mm F1.4 AS UMC). The error in energy reading of the electron spectrometer was determined by electron beam size and pointing and varied depending on energy between (+6, -3.2) MeV for 175 MeV and (+35, -15.7) MeV for 400 MeV. The energy calibration of the electron spectrometer was done using the G4beamline code[31].

After energy calibration, the peak energy (the part of the energy spectrum that contains most of the charge) was determined by finding the maximum of counts from the image and reading the corresponding energy. The data taken in the same experimental conditions were then averaged and the mean value of energy with its corresponding mean standard error was obtained. For each dataset, the maximum value of energy was extracted.

**Particle in cell simulations**

To understand the underlying physics multiple PIC quasi-3D simulations with OSIRIS[32,33] were performed. Here the results are discussed for three density profiles plotted in Fig. 2 (a), where in case i) The plasma density rises to $n_0$ in plasma scale length 900 $c\backslash\omega_p$, stay flat for 800 $c\backslash\omega_p$, and drops down to 0 in next 900 $c\backslash\omega_p$ plasma length. In case ii) after reaching the density $n_0$ in plasma scale length 900 $c\backslash\omega_p$, the electron density further grows to 1.4 $n_0$ in next 800 $c\backslash\omega_p$, and in the case iii) the density profile is similar to the case i, with flat density at 1.4 $n_0$. Rest of the laser parameters and simulation setup is identical in all the simulations, and are discussed below.

We initialize a simulation box, which moves with the speed of light c, has dimensions of $36c/\omega_p \times 50c/\omega_p$ and is divided into $3600 \times 500$ cells, along z and r direction, respectively, with 32 particles per cell. The laser and plasma parameters were chosen by taking into account the experimental conditions, and are as follows: laser peak normalized vector potential $a_0 = 1.2$, gaussian transverse profile with $42\mu m$ full-width half-maximum spot size and 30fs full-width half-maximum pulse length, and $n_0 = 2.0 \times 10^{19}\ cm^{-3}$.

## RESULTS

The experiment has been done using the setup described in Methods. As a reference, we use the dataset recorded with the density profile P3 shown in Fig. 1 in Supplementary Material. The density profiles are obtained from computational fluid dynamics simulations (see Methods). Various gas density profiles were generated by adjusting the tilt angle of the nozzle from 0° to 30°. A further change in the inlet pressure and the interaction position above the nozzle was required to stabilize the electron beam in terms of pointing, charge and divergence. For each of the density profiles shown in Fig. 2 Supplementary Material (experimental conditions summarized in Table 2 in Supplementary Material), we recorded a dataset and the results were averaged. For clarity, Fig 1c only shows two density profiles corresponding to 0° and 10° tilting angle. The tilting of the nozzle affects the slope of the density profile ramps and transforms the flat top profile region into a ramp. The mean energy and mean divergence are shown in Fig 1a and Fig 1d respectively. The electron beam average peak energy for the gas profile with 0° tilting (red colour dashed profile in Fig. 1c) was 174.8 ± 1.33 MeV and increased to 226.2 ± 9.49 MeV and 262 ± 9.73 when the density profiles were generated with tilting angles of 5° and 10°. Interestingly, the gas density profile for 10° tilt angle (black curve in Fig. 1c) has almost the same length as the density profile generated by 0°, with a 6.6% increase in peak density. Therefore we can assume that the increase in electron energy cannot be due to an increase in plasma length but to changes in the density profile slope. For further tilting to 20° and 30° the electron energy reached a plateau and slightly decreased to 243.2 ± 7.93 MeV and 238.9 ± 8.29 MeV. The shots with maximum peak electron energy for each tilting angle are shown in Fig. 1b. The peak energy increased from 200 MeV for 0° up to 363 MeV for 10° tilting angle, values significantly higher than the mean energies obtained for each dataset, highlighting the effect of fluctuations in the laser parameters. Due to the shot-to-shot fluctuations of the laser focal plane, the spot diameter (not shown here) had between ± 10.5% and ± 5.4% maximum size fluctuation along the horizontal and the vertical axis respectively for 40 consecutive laser shots. Also, important parameters of the laser pulse, such as spectral phase or temporal pulse shape are not known on a shot-to-shot basis, due to experimental diagnostic constraints. In our experiment injection and acceleration happened mainly in the plasma up-ramp thus the initial laser conditions were critical for the outcome. The acceleration process could be initiated in different points in the ramp where the plasma conditions were very different, consequently affecting the evolution of the wakefield. An increased shot-to-shot stability of the laser should yield an electron beam which presents very small shot-to-shot fluctuations of its parameters.

Due to the large acceptance angle of the electron spectrometer (± 9.3 mrad at the entrance and 6.3 mrad at exit), shot-to-shot electron beam pointing vertical stability (on average between -1 mrad and +8 mrad) and electron beam size we could not deconvolve the energy spectrum to recover the real energy spread. Therefore this parameter was not discussed in the present work but related details can be found in another study[34]. We note that the acceptance angle of the spectrometer was much smaller than the divergence of the electron beam and it limited the transport efficiency especially at lower energies where the divergence was significant, thus any information about real beam size or charge was lost after the electron spectrometer.

The mean divergence of the electron beam followed closely the trend of the relative charge (see Fig. 1a red curve) and showed a significant decrease along both vertical (DivY) and horizontal (DivX) axis: from DivX = 58.9 ± 0.45 mrad and DivY = 35.2 ± 0.27 mrad (for the density profile corresponding to 0°) dropped to DivX = 12.6 ± 1.17 mrad and DivY = 8.2 ± 0.69 mrad (for the density profile corresponding to 5° and 10° respectively). The reduction of the divergence is assumed to be due to selective injection of electrons into the bubble which occurs mainly near the axis[21] thus the accelerated electrons have a lower transverse momentum (i.e. lower emittance).

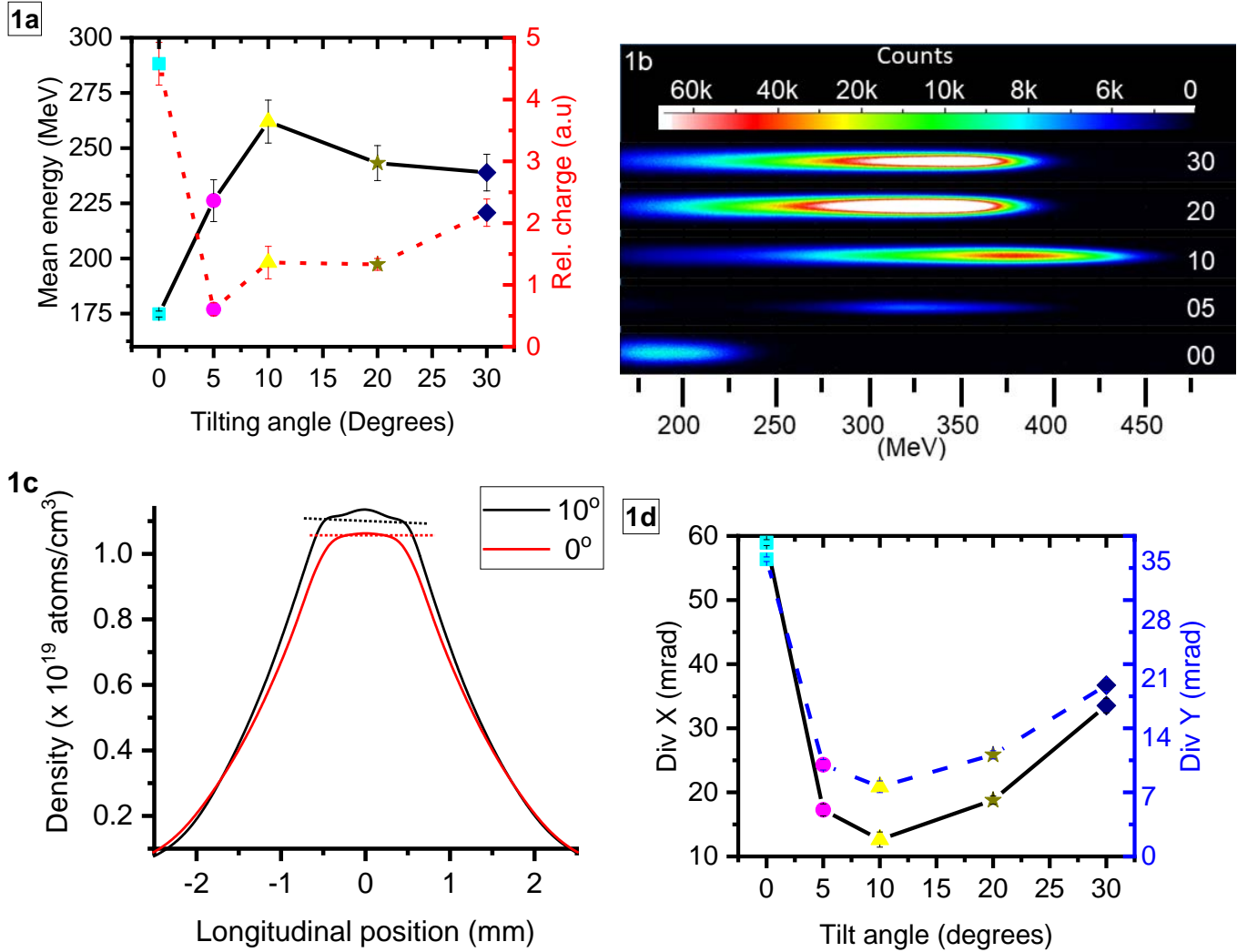

*Figure 2 Fig. 1c shows the gas density profiles corresponding to datasets obtained with 10° and 0° tilting angle. The laser propagates from the right to left. The dotted lines are drawn just to guide the eye and point out the small difference between the top part of the density profiles. The density profiles were obtained by tilting the nozzle, adjusting the interaction point in the vertical direction and changing the inlet pressure. In Fig. 1a is shown the evolution of the mean electron energy for various tilting angles. The error bars represent the standard error of the mean. From each dataset, one shot at maximum energy was extracted and shown as a false colour image in Fig. 1b. The mean electron beam divergence is shown in Fig. 1d and follows closely the variation of the mean electron energy. Each dataset is colour-coded depending on the density profile (density profiles shown in Supplementary Material): cyan for the profile at 0°, magenta for 5°, yellow for 10°, dark yellow for 20° and navy for 30°.*

## DISCUSSION

In order to check if relatively small changes in the plasma density ramp can determine changes in the electron beam parameters we did full scale three dimensional PIC simulation with OSIRIS[32,33] using 3 types of density profiles, shown in Fig 2a: one trapezoidal profile with peak density $1\times10^{19}$ electrons/cm$^3$ (case *i*), one trapezoidal profile with peak density $1.4\times10^{19}$ electrons/cm$^3$ (case *iii*) and one profiles with 3 ramps (case *ii*). For the case *i* and *iii*, the peak energies were very similar. Even though due to higher density, in case *iii* the self-injection occurred at early stage and acceleration field was stronger, faster laser depletion limited the peak energy for the accelerated electrons to 250 MeV. On the other hand in case i peak energy reached 275 MeV. By introducing another density up-ramp in the previously flat part, i.e. in case *ii*, the dephasing of the electron bunch was modified to reach peak energy 315 MeV. Tailored plasma density can provide electron energy enhancement due to the combined effect of phase-locking and enhanced acceleration field, without drastically reducing the acceleration length due to the laser-etching (depletion). It is worth mentioning here that the injection for the case (i) and (ii) occur at the same position, however at a later stage as compare to the case (iii) where the laser effectively sees a higher density as well as density gradients.

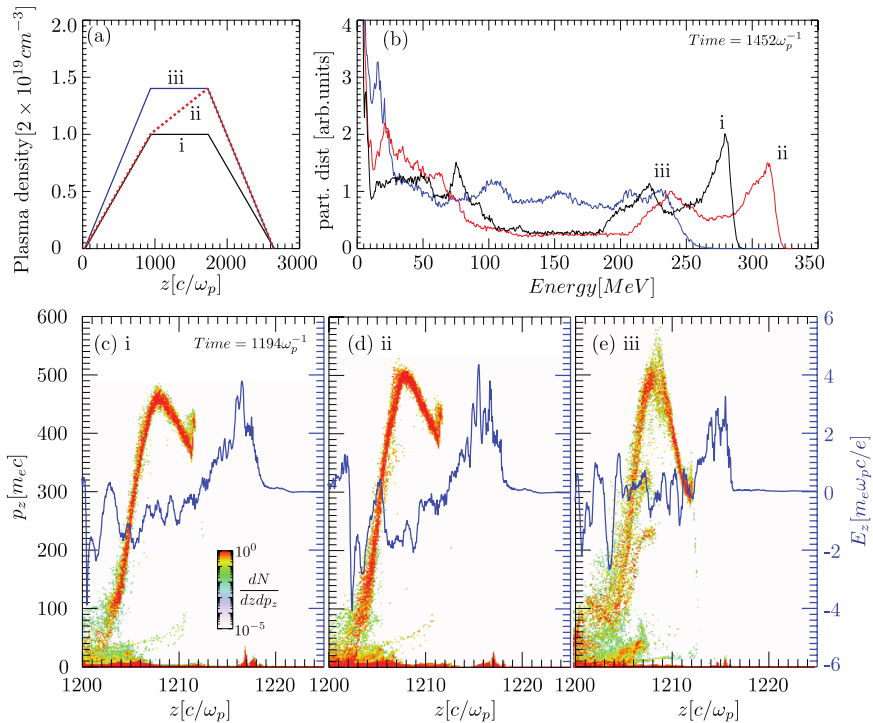

Figure 3 Quasi-3D PIC simulation to show the of density ramp on the LWFA: Figure (a) highlights the initial density profiles used to simulate three cases, and (b) compares the electron energy spectrum for the three cases at time $t = 1452\ \omega_p^{-1}$. Figure (c), (d), and (e) show the longitudinal phase space and electric field lineout at the axis (blue line) for the three cases at time $t = 1194\ \omega_p^{-1}$.

In our experiment with laser wakefield acceleration in up-ramp plasma densities, the control of the electron beam energy was achieved while the laser power was kept constant. The same level of energy tuning can be obtained, theoretically, in a homogeneous plasma by increasing the laser power more than 5 times[35], a result that proves that the method presented by us can improve the efficiency of the acceleration process. Although we varied the density profile quite significantly with the nozzle tilted at 0º (see Supplementary Material), negligible changes could be observed in the parameters of the electron beam, a result that supports our hypothesis that the phase of the wakefield can be manipulated by upward plasma densities. Particle-in-cell simulations showed that the presence of an up-ramp density instead of a flat top profile can boost the electron energy although not to the levels seen in the experiment. At this point, we cannot be certain why better matching between simulations and experiments cannot be achieved. We will investigate the possibility to include in future simulations real laser parameters such as spectral phase, wavefront shape, etc. which, due to a strong nonlinear evolution, could further boost and enhance the parameters of the accelerated electron beam.

Our result can be especially useful for LWFA with the new generation of kHz lasers. Due to technological constraints, the kHz laser energy is limited to a few 10's of mJ, therefore, a method to increase the electron energy for a given laser energy would greatly enhance the development of these type of LWFA accelerators. Scanning systematically and automatically the density profiles, (e.g. through learning algorithms) one can build a map of electron energies for various sets of parameters thus creating an energy tunable source of electrons.

The increase of the energy of electrons will have a crucial impact in realizing future laser-electron scattering experiments. In fundamental studies, it can allow a systematic access of distinct radiation reaction regimes[36,37] and in application studies our results can lead to tunable gamma radiation sources. The effects are highest for Inverse Compton Scattering experiments[38], where the energy of gamma-ray photons depends quadratically on the electron energy and the divergence is inversely proportional to it. Improvements of these parameters will have far reaching consequences for the applicability of new radiation sources in non-destructive imaging and nuclear activation[15].

Last but not least, many of the plasma-based accelerators nowadays use uniform density profiles but our results suggest that upward plasma ramps should be used instead if one wants to maximize the electron energy output of the accelerator and improve the quality of the electron beam.

**Contributions**
C.A conceived the idea, designed the setup, participated in the experiment, performed the Ansys Fluent simulations and extracted the gas density profiles, run data analysis and wrote the manuscript, B.J.Y offered technical support for the experimental setup design, K.H.O, C.I.H, J.H.J, S.H.K, J.H.S participated in the experiment, E.B wrote the code used for data acquisition and processing, V.B.P performed the PIC simulations, Y.H.Y, J.H.S and S.K.L supported the laser maintenance, H.T.K, M.H.C, B.M.H and C.H.N offered support for manuscript writing, correction and revision. All the authors reviewed the manuscript.

## Acknowledgements

The authors thank the OSIRIS consortium at UCLA and IST for providing access to OSIRIS 3.0 framework.

This work has been supported by the Institute for Basic Science of Korea under IBS-R012-D1.

E.B. acknowledges funding from U.K. EPSRC (EP/J018171/1, EP/N028694/1)

## Competing financial interests

The authors declare that they have no competing financial interests.

### Data availability

The raw data is available from the corresponding author on reasonable request.

## References

1. Hora, H. Self-focusing of laser beams in a plasma by ponderomotive forces. *Zeitschrift für Phys.* **226,** 156–159 (1969).
2. Bituk, D. R. & Fedorov, M. V. Relativistic ponderomotive forces. *J. Exp. Theor. Phys.* **89,** 640–646 (1999).
3. Tajima, T. & Dawson, J. M. Laser Electron Accelerator. *Phys. Rev. Lett.* **43,** 267–270 (1979).
4. Pukhov, A. & Meyer-ter-Vehn, J. Laser wake field acceleration: The highly non-linear broken-wave regime. *Appl. Phys. B Lasers Opt.* **74,** 355–361 (2002).
5. Lu, W., Huang, C., Zhou, M., Mori, W. B. & Katsouleas, T. Nonlinear theory for relativistic plasma wakefields in the blowout regime. *Phys. Rev. Lett.* **96,** 165002 (2006).
6. Lu, W. *et al.* Generating multi-GeV electron bunches using single stage laser wakefield acceleration in a 3D nonlinear regime. *Phys. Rev. Spec. Top. - Accel. Beams* **10,** 061301 (2007).
7. Naumova, N. *et al.* Attosecond electron bunches. *Phys. Rev. Lett.* **93,** 195003 (2004).
8. Wang, W. T. *et al.* High-Brightness High-Energy Electron Beams from a Laser Wakefield Accelerator via Energy Chirp Control. *Phys. Rev. Lett.* **117,** 124801 (2016).
9. Islam, M. R. *et al.* Near-threshold electron injection in the laser-plasma wakefield accelerator leading to femtosecond bunches. *New J. Phys.* **17,** 093033 (2015).
10. Brunetti, E. *et al.* Low Emittance, High Brilliance Relativistic Electron Beams from a Laser-Plasma Accelerator. *Phys. Rev. Lett.* **105,** 215007 (2010).
11. Khachatryan, A. G., Van Goor, F. A. & Boller, K. J. Coherent and incoherent radiation from a channel-guided laser wakefield accelerator. *New J. Phys.* **10,** 084043 (2008).
12. Hooker, S. M. *et al.* Multi-pulse laser wakefield acceleration: A new route to efficient, high-repetition-rate plasma accelerators and high flux radiation sources. *J. Phys. B At. Mol. Opt. Phys.* **47,** 234003 (2014).
13. Schlenvoigt, H. P. *et al.* A compact synchrotron radiation source driven by a laser-plasma wakefield accelerator. *Nat. Phys.* **4,** 130–133 (2008).
14. Döpp, A. *et al.* A bremsstrahlung gamma-ray source based on stable ionization injection of electrons into a laser wakefield accelerator. *Nucl. Instruments Methods Phys. Res. Sect. A Accel. Spectrometers, Detect. Assoc. Equip.* **830,** 515–519 (2016).
15. Albert, F. & Thomas, A. G. R. Applications of laser wakefield accelerator-based light sources. *Plasma Phys. Control. Fusion* **58,** 103001 (2016).
16. Esarey, E., Schroeder, C. B. & Leemans, W. P. Physics of laser-driven plasma-based electron accelerators. *Rev. Mod. Phys.* **81,** 1229–1285 (2009).
17. Katsouleas, T. Physical mechanisms in the plasma wake-field accelerator. *Phys. Rev. A* **33,** 2056–2064 (1986).
18. Wen, M. *et al.* Controlled electron acceleration in the bubble regime by optimizing plasma density. *New J. Phys.* **12,** 045010 (2010).
19. Pukhov, A. & Kostyukov, I. Control of laser-wakefield acceleration by the plasma-density profile. *Phys. Rev. E - Stat. Nonlinear, Soft Matter Phys.* **77,** 025401 (2008).
20. Rittershofer, W., Schroeder, C. B., Esarey, E., Grüner, F. J. & Leemans, W. P. Tapered plasma channels to phase-lock accelerating and focusing forces in laser-plasma accelerators. *Phys. Plasmas* **17,** 063104 (2010).
21. Yu, Q. *et al.* Electron self-injection into the phase of a wake excited by a driver laser in a nonuniform density target. *Phys. Plasmas* **22,** 073107 (2015).
22. Yoon, S. J., Palastro, J. P. & Milchberg, H. M. Quasi-phase-matched laser wakefield acceleration. *Phys. Rev. Lett.* **112,** 134803 (2014).
23. Geun Ju, K. & Seung Hoon, Y. Energy Enhancement Using Upward Density Ramp in Laser Wakefield

24. Strickland, D. & Mourou, G. Compression of amplified chirped optical pulses. *Opt. Commun.* **56,** 219–221 (1985).
25. Fluent, a. ANSYS Fluent 12.0 user's guide. *Ansys Inc* **15317,** 1–2498 (2009).
26. Anderson, J. D. *Computational fluid dynamics : the basics with applications. McGraw-Hill series in mechanical engineering McGraw-Hill series in aeronautical and aerospace engineering McGraw-Hill international editions. Mechanical engineering series* **Internatio,** (1995).
27. Schmid, K. & Veisz, L. Supersonic gas jets for laser-plasma experiments. *Review of Scientific Instruments* **83,** 53304 (2012).
28. Esarey, E., Ride, S. K. & Sprangle, P. Nonlinear Thomson scattering of intense laser pulses from beams and plasmas. *Phys. Rev. E* **48,** 3003–3021 (1993).
29. Chen, S. Y., Maksimchuk, A. & Umstadter, D. Experimental observation of relativistic nonlinear Thomson scattering. *Nature* **396,** 653–655 (1998).
30. Clayton, C. E. *et al.* Plasma wave generation in a self-focused channel of a relativistically intense laser pulse. *Phys. Rev. Lett.* **81,** 100–103 (1998).
31. Roberts, T. J. & Kaplan, D. M. G4beamline simulation program for matter-dominated beamlines. in *Proceedings of the IEEE Particle Accelerator Conference* 3468–3470 (2007). doi:10.1109/PAC.2007.4440461
32. Fonseca, R. A. *et al.* OSIRIS: A Three-Dimensional, Fully Relativistic Particle in Cell Code for Modeling Plasma Based Accelerators. in *International Conference on Computational Science* (2002).
33. Fonseca, R. A. *et al.* One-to-one direct modeling of experiments and astrophysical scenarios: pushing the envelope on kinetic plasma simulations. *Plasma Phys. Control. Fusion* **50,** 124034 (2008).
34. Aniculaesei, C. Department of Physics Experimental Studies of Laser Plasma Wakefield Acceleration. (University of Strathclyde, 2015).
35. Lu, W. *et al.* Designing LWFA in the blowout regime. in *2007 IEEE Particle Accelerator Conference (PAC)* 3050–3051 (2007). doi:10.1109/PAC.2007.4440664
36. Poder, K. *et al.* Experimental Signatures of the Quantum Nature of Radiation Reaction in the Field of an Ultraintense Laser. *Phys. Rev. X* **8,** (2018).
37. Cole, J. M. *et al.* Experimental Evidence of Radiation Reaction in the Collision of a High-Intensity Laser Pulse with a Laser-Wakefield Accelerated Electron Beam. *Phys. Rev. X* **8,** (2018).
38. Chen, S. *et al.* MeV-energy X rays from inverse compton scattering with laser-wakefield accelerated electrons. *Phys. Rev. Lett.* **110,** (2013).

# SUPPLEMENTARY MATERIAL

**Experimental results with a straight nozzle**

The gas nozzle was set in straight position (gas flow direction perpendicular to the laser axis) and the laser focal plane fixed longitudinally at 1.5 mm from the centre of the nozzle. Helium gas was used. Various gas density profiles, as seen by the laser, were generated by changing the inlet pressure and the interaction point relative to the exit of the nozzle (the height).

The gas density profiles are shown in Fig. 1a, each colour corresponding to a dataset recorded in the same experimental conditions. The results from each dataset are averaged. The specific experimental conditions for each dataset are summarized in Table 1. The results for the mean electron energy shown in Fig. 1.b present a very small variation, around 5% from 174.8 ± 1.33 MeV (yellow data set in Fig. 1b) to 183.7 ± 5.97 MeV (blue dataset in Fig. 1b) even though the gas density has been varied more than 50% from $0.88 \times 10^{19}$ atoms/cm$^3$ (black curve P1) to $1.37 \times 10^{19}$ atoms/cm$^3$ (blue curve P4). The error bar represents the standard error of the mean. The mean divergence remains quasi-constant, ~58 mrad along the horizontal axis and ~35 mrad along the vertical axis. The stability of the electron beam energy is the only one that suffers changes, especially at higher gas densities where, for example, it shows a 4.4 times increase of the mean energy standard deviation between dataset obtained with the profile P3 compared with the dataset obtained the profile P4.

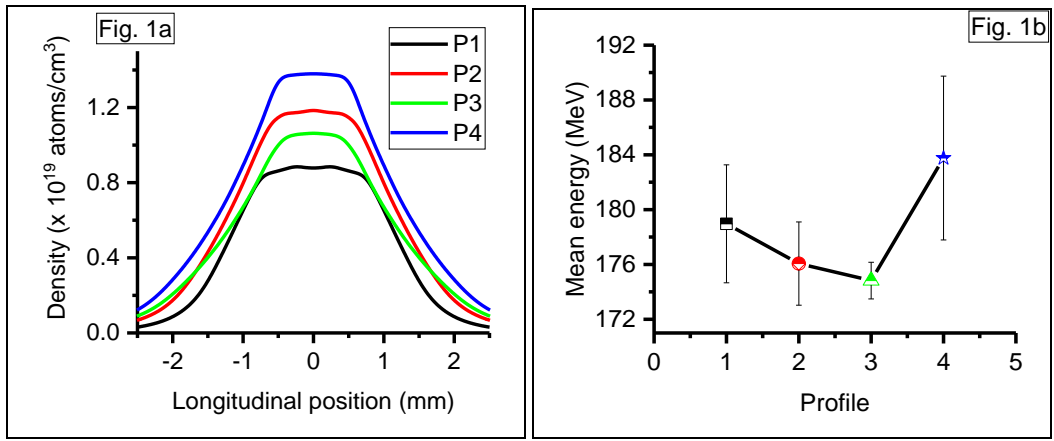

*Fig. 1 The experimental results obtained with the straight nozzle. In Fig.1a are shown the gas density profiles and in Fig.1b are shown the corresponding mean energy for each dataset. The peak density has been adjusted more than 50%, from $0.88 \times 10^{19}$ atoms/cm$^3$ (black curve) to $1.37 \times 10^{19}$ atoms/cm$^3$ (blue curve) and the mean electron energy showed a negligible variation of 5%. The divergence remains quasi-constant, around 58 mrad along the horizontal axis and around 35 mrad along the vertical axis. The laser propagates from right to left.*

| Profile | No of shots | Height (mm) | Inlet pressure (bar) | Peak density (×10$^{19}$ atoms/cm$^3$) |
|---|---|---|---|---|
| 1 (Black) | 12 | 3 | 10 | 0.88 |
| 2 (Red) | 13 | 4 | 15 | 1.18 |
| 3 (Green) | 12 | 5 | 15 | 1.06 |
| 4 (Blue) | 14 | 5 | 20 | 1.37 |

*Table 1 contains the experimental conditions for the straight nozzle case.*

## SUMMARIZED EXPERIMENTAL RESULTS FOR THE TILTED NOZZLE CASE

In Table 2 the first ramp is the ramp that interacts first with the laser (the laser propagates from right to left in Fig. 1c). The 1$^{st}$ ramp starts where the gas density is $1 \times 10^{18}$ atoms/cm$^3$ and stops at the beginning of the second ramp. The 2$^{nd}$ ramp stops where the gas density reaches its maximum and the 3$^{rd}$ ramp starts at the end of 2$^{nd}$ ramp end ends where the gas density drops to $1 \times 10^{18}$ atoms/cm$^3$.

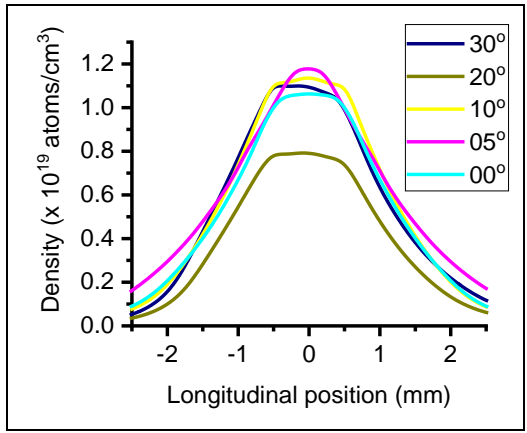

*Fig. 2 shows the gas density profile corresponding to each dataset with the laser propagating from the right side towards the left side. The density profiles were obtained by tilting the nozzle, adjusting the interaction point in the vertical direction and changing the inlet pressure, matching the parameters in Table 2.*

| Tilt angle α (°) | No of shots | Height (mm) | Inlet Pressure (bar) | Peak density (×10$^{19}$ atoms/cm$^3$) | 1$^{st}$ ramp length (mm) | 2$^{nd}$ ramp length (mm) | 3$^{rd}$ ramp length (mm) | Electron mean peak energy (MeV) |
|---|---|---|---|---|---|---|---|---|
| 30 (navy) | 20 | 4 | 15 | 1.09 | 2.184 | 0.869 | 1.744 | 238.9 |
| 20 (dark yellow) | 23 | 3 | 10 | 0.79 | 1.741 | 0.942 | 1.492 | 243.2 |
| 10 (yellow) | 21 | 3 | 15 | 1.13 | 2.016 | 0.801 | 1.946 | 262 |
| 5 (magenta) | 15 | 5 | 20 | 1.17 | 2.9 | 0 | 2.9 | 226.2 |
| 0 (cyan) | 12 | 5 | 15 | 1.06 | 2.029 | 0.806 | 2.029 | 174.8 |

*Table 2 The experimental conditions for the tilted nozzle case*